\documentclass[%
 reprint,
 amsmath,amssymb,
 aps,
]{revtex4-2}

\usepackage{graphicx}% Include figure files
\usepackage{dcolumn}% Align table columns on decimal point
\usepackage{bm}% bold math
\usepackage{bbding}
\usepackage{multirow}
\usepackage{hyperref}
\usepackage{subcaption} 

\begin{document}

\preprint{APS/123-QED}

\title{Monte Carlo study of KDAR $\nu_{\mu}$ charged-current scattering on carbon}

\author{R K Pradhan\textsuperscript{1}}%
 \email{kumarriteshpradhan@gmail.com}

\author{R Lalnuntluanga\textsuperscript{2}}
\email{tluangaralte.phy@gmail.com}

\author{A Giri\textsuperscript{1}}%
 \email{giria@phy.iith.ac.in}
 
\affiliation{%
\textsuperscript{1}
 \textit{Indian Institute of Technology Hyderabad, Hyderabad, 502284, Telangana, India}
}%

\affiliation{\textsuperscript{2}
\textit{Tel Aviv University, Tel Aviv 69978, Israel}
}

\begin{abstract}
The neutrino energy reconstruction is crucial for reducing systematic uncertainties in neutrino oscillation experiments and improving cross-section measurements. Kaon-Decays-At-Rest (KDAR) neutrinos provide a unique opportunity to probe neutrino-nucleus interactions at low energies with a precisely known energy of 235.5 MeV. This work presents Monte Carlo predictions of missing energy due to nuclear effects in the KDAR $\nu_\mu$ charged-current scattering on the carbon nucleus for JSNS$^2$. The predictions from GENIE, NuWro, and GiBUU show a significant impact of nuclear and final state interaction (FSI) models, as well as nucleon removal energy, on missing energy reconstruction. Tuning the hA FSI model improves the predictions from GENIE. It is also observed that the nucleon elastic component of FSI leads to better predictions in NuWro compared to GENIE. NuWro shows the best agreement with data among the generators considered; however, the observed discrepancies with data across all generators highlight the limitations of current theoretical descriptions in the low-energy regime and the need for improvements in Monte Carlo modeling.
\end{abstract}

\maketitle

%\tableofcontents

\section{\label{intro}Introduction}

In the precision measurement era, the uncertainty in neutrino-nucleus interaction models needs to be reduced to achieve the goals of the current and future accelerator-based neutrino experiments \cite{NuSTEC:2017hzk}. For a better understanding of neutrino-nucleus interactions, experiments such as DUNE \cite{DUNE:2020ypp}, Hyper-Kamiokande \cite{Hyper-Kamiokande:2018ofw}, and JUNO \cite{JUNO:2015zny} use different target materials and provide a wide range of neutrino fluxes. Statistical requirements require the use of heavy nuclear targets, which introduce systematic uncertainties in cross-sectional measurements due to the complexity of the nuclear medium, ultimately affecting the measurements of the oscillation parameters \cite{NOvA:2019cyt,T2K:2019bcf}. The complex nuclear medium gives rise to various nuclear effects, including initial-state and final-state interactions, which affect the final state of neutrino interactions. The neutrino-nucleon scattering can be understood by the impulse approximation (IA) \cite{Benhar:2005dj}, which is a two-step process. The first step is the primary interaction of a neutrino on a bound nucleon or a nucleon pair which is affected by the the initial state nuclear effect such as Fermi motion \cite{Pradhan:2024gqv}, Pauli blocking \cite{Bodek:2021trq}, the removal energy due to the nuclear potential, and multi-nucleon interaction \cite{Sobczyk:2012ms} that describe the ground state of the nucleus. The second step in the IA picture is the hadron rescatterings known as Final state interaction (FSI) \cite{Dytman:2009zz}. In FSI, the hadrons produced inside the nucleus undergo rescattering, before coming out of the nucleus, through various hadronic interactions such as pion production(absorption), (in)elastic, and charge exchange, leading to the misidentification of the processes. A quasi-elastic (QE) scattering event can be misidentified as non-QE due to pion production in FSI, while a resonance (RES) scattering event can be misidentified as QE due to pion absorption. Hence, a proper event selection is required to reconstruct a specific interaction. For example, an event selection study to reconstruct the neutrino energy for charge current quasi-elastic (CCQE) scattering can be found in Ref. \cite{Lalnuntluanga:2024lti}. 

Neutrino beams are produced by the decay of mesons, mostly pions and kaons, generated in proton-nucleus scattering at accelerator facilities. Neutrinos produced via decay-in-flight have a broad energy spectrum, requiring experiments to reconstruct the incident neutrino energy on an event-by-event basis which is a model-dependent analysis. Kaon-decays-at-rest (KDAR) provides mono-energetic muon neutrinos (235.5 MeV) via $K^+ \rightarrow \mu^+\nu_{\mu}$ with a branching ratio of 63.6\%  \cite{ParticleDataGroup:2022pth}. The KDAR $\nu_{\mu}$ beam is an excellent tool for studying neutrino-nucleus interaction and determining weak interaction parameters because of their known energy without the complications of the broad energy spectrum. In addition, KDAR neutrinos are also useful for the sterile neutrino study \cite{LSND:2001aii,MiniBooNE:2018dus} as well as for exploring the interactions of the core collapsing supernova \cite{Kusakabe:2019znq,Ko:2022uqv}. Several current and planned experiments have begun studying KDAR neutrinos. The first observation of KDAR $\nu_{\mu}$ interactions with carbon was reported by MiniBooNE \cite{MiniBooNE:2018dus}, and more recently, the J-PARC Sterile Neutrino Search at the J-PARC Spallation Neutron Source (JSNS$^2$) measured KDAR $\nu_{\mu}$ scattering on carbon \cite{JSNS2:2024bzf}.

In this work, we reconstructed the missing energy caused by nuclear effects in the KDAR $\nu_{\mu}$ scattering on carbon using various Monte Carlo generator models in GENIE \cite{Andreopoulos:2015wxa}, NuWro \cite{Juszczak:2005zs}, and GiBUU \cite{Buss:2011mx}. The Monte Carlo predictions were then compared with the JSNS$^2$ measurement \cite{JSNS2:2024bzf}. In addition to nuclear effects, missing energy also arises from undetected neutrons and hadrons that escape the detector due to its limitations. This work presents reconstructed missing energy using different nuclear and FSI models. The hA FSI model is tuned on the data and shows improved predictions from GENIE. However, a significant discrepancy between existing models and measurements indicates the limitation of current generators.

This article is organized as follows: In Sec. \ref{sec2}, we present the reconstruction mechanism of the missing energy. The simulation specifications, including Monte-Carlo generators, are described in Sec. \ref{sec3}. The results and conclusion from the analysis are discussed in Secs. \ref{sec4} and \ref{sec5} respectively. 

\section{\label{sec2}Missing Energy Reconstruction}

The missing energy is the difference between the energy transferred to the nucleus ($\omega$) and the kinetic energy of the hadrons measured by the detector. Since KDAR $\nu_{\mu}$'s are mono-energetic ($E_{\nu_\mu}$ = 235.5 MeV), the energy transferred can be calculated from the kinetic energy of the muon ($T_{\mu}$) as follows,
\begin{equation*}
    \omega = E_{\nu_{\mu}} - E_{\mu} = E_{\nu_{\mu}} - m_{\mu}-T_{\mu}
\end{equation*}
where $m_{\mu}$ is the mass of the muon (105.7 MeV). The missing energy is defined as,
\begin{equation*}
    E_{miss} = \omega - \sum T_h  = E_{\nu_{\mu}} - m_{\mu}-T_{\mu} - \sum T_h 
\end{equation*}
\begin{equation}
\label{eq1}
\begin{aligned}
    E_{miss} &= E_{\nu_{\mu}} - m_{\mu} - E_{viss}\\
             &= 235.5 - 105.7 - E_{viss} \\
             &= 129.8 \text{ MeV} - E_{viss}
    \end{aligned}
\end{equation}

Here $E_{viss} =T_{\mu} +\sum {T_h} $, is the kinetic energy of the final state particles detected, and $T_h$ is the kinetic energy of the hadrons. Since KDAR neutrinos' energy is in the MeV scale, we can naively consider that they interact quasi-elastically via $\nu_\mu n \rightarrow \mu^- p$ and one can calculate $E_{viss}$ using the calorimetric method as,
\begin{equation*}
    \begin{aligned}
        E_{viss} &= E_{\nu_\mu} - m_{\mu} + (m_n - m_p) \\
                 &= 235.5 - 105.7 + 1.3 \\
                 &= 131.2 \text{ MeV}
    \end{aligned}
\end{equation*}
However, the above expectation is modified significantly due to various nuclear effects described in \ref{intro}. In this work, we calculated the shape-only differential cross-section, $\frac{1}{\sigma} \frac{d\sigma}{dE_m}$ and compared it with the unfolded shape-only cross-section data obtained from Fig. 4 in Ref. \cite{JSNS2:2024bzf}.

\section{\label{sec3}Simulation Framework}

This work uses GENIE, NuWro, and GiBUU to calculate neutrino-nucleus cross-section and simulate mono-energetic $\nu_\mu$ interactions on Carbon nucleus, considering the charged current (CC) Quasi-elastic (QE), Resonance (RES), Deep inelastic (DIS), Meson exchange (MEC),
and Coherent (COH) channels. Since the CCQE channel mostly dominates KDAR $\nu_\mu$ CC scattering, this analysis applies a CCQE event selection. Selected events are defined as those with a final state consisting of $1\mu^- 1p\,0\pi \,Xn$ \cite{Lalnuntluanga:2024lti}.\\
\\
\textbf{GENIE:} 1 million KDAR $\nu_\mu$ events on Carbon are simulated using GENIE v3.06.00 with the tune G18\_10a\_02\_11a. GENIE describes the ground state of the nucleus using the Fermi Gas model and spectral function. In this simulation, the models for the ground state of the nucleus used are the Local Fermi Gas model (LFG) \cite{Bodek:1981wr}, Correlated Fermi Gas model (CFG) \cite{CLAS:2005ola}, Spectral Function (SF) \cite{Benhar:1994hw}, Spectral function-like-LFG, Spectral function-like-CFG, and continuum random phase approximation (CRPA) \cite{Nikolakopoulos:2020alk}. GENIE considers the nucleon-nucleon correlation quantified by 20\% of the short-range correlation (SRC) fraction \cite{Cruz-Torres:2017sjy}. The SF-like-LFG is a modification of the LFG model, introducing a momentum-dependent removal energy for a more realistic treatment of nucleon binding. The removal energy becomes a shift in four-momentum transferred ($Q^2$), and the lowest removal energy of the spectrum corresponds to the binding energy of the p-shell of the nucleus. The SF-like-CFG is a similar model to SF-like-LFG, considering the correlation between nucleons. The CRPA model is important for the low-energy regions and the lighter nuclei, such as carbon and oxygen. In case of the CRPA framework, the nucleus is described using the Hartree-Fock (HF) mean-field model with a Skyrme-type nucleon-nucleon interaction. This provides a realistic description of the nuclear ground state. The nucleon removal energy is set to 15 MeV. The Valencia Model \cite{Gran:2013kda} is used to simulate CCQE scattering. The Valencia QE model is based on the Random Phase Approximation (RPA) \cite{Co:2023ktj} and considers Coulomb correction effects. The dipole approximation is used for the QE axial form factor, taking its value as -1.267 at $Q^2$ = 0. The axial ($M_A$) and vector ($M_V$) masses for QE are set to 0.96 and 0.84 GeV, respectively. The BBA07 elastic form factor model \cite{Bodek:2007ym} is used to calculate the CCQE cross-section. The CCMEC scattering is simulated by the Valencia MEC model \cite{Nieves:2016sma} with a maximum momentum transferred of 1.2 GeV. Since CCQE and CCMEC are the dominant channels in KDAR $\nu_\mu$ scattering, the default models are used to simulate other channels. For the CCRES channel, the Berger-Sehgal (BS) model \cite{Berger:2007rq} is used for cross-section calculations, and all 17 resonances are simulated with an axial mass of 1.065 GeV. The CCDIS events are simulated using the Bodek-Yang model \cite{Bodek:2002vp} with KNO scaling \cite{Koba:1972ng}. The invariant hadronic mass threshold for the RES-DIS joining scheme is taken as 1.9 GeV. The FSI models used in this simulation are hA, hN, and Geant4 \cite{GENIE:2021npt}.\\ 
\\
\textbf{NuWro:} NuWro v21.09.2 is used to simulate KDAR $\nu_\mu$ CC scattering and calculate cross-sections. The initial nuclear state is modeled using LFG, Relativistic Fermi Gas Model (RFG), and SF in NuWro. The SF model includes the effect of nucleon short-range correlated (SRC) pairs that cause a large tail in the target nucleon momentum distribution. The binding energy for the carbon nucleus is taken as 92.16 MeV. The CCQE scattering is simulated using the Llewellyn-Smith (LS) model \cite{LlewellynSmith:1971uhs}. The axial and vector form factors used for CCQE are dipole approximation and BBBA05 \cite{Bradford:2006yz} respectively with an axial mass of 0.96 GeV. NuWro uses the same models as GENIE to simulate CCMEC scattering. NuWro uses the Rein-Sehgal (RS) model \cite{Rein:1980wg} for $\Delta$(1232) resonance production with the form factor model by Paschos and Lalakulich \cite{Lalakulich:2005cs}. The DIS interaction is simulated using the BS model and hadronic final states are generated with PYTHIA \cite{Sjostrand:2006za}. The threshold for RES-DIS transition is set at 1.8 GeV. NuWro uses the BS model to simulate COH scattering including RS model correction for CC coherent single-pion production. The FSI effects are modeled by the cascade model based on the Metropolis algorithm \cite{Metropolis:1958sb}. NuWro employs the Formation Zone (FZ) effects \cite{Golan:2012wx} for the hadrons produced in the nuclear medium. The FZ effect provides a time frame for hadrons to form before they interact with the nuclear medium. The models used for the FZ effect are: the coherence length model for QE \cite{Battistoni:2009zzb}, the $\Delta$ lifetime-based model for RES \cite{Golan:2012wx}, and the model based on hadron-hadron and hadron-nucleus collision for DIS \cite{Ranft:1988kc}.\\
\\
\textbf{GiBUU:} The recent release of GiBUU 2025 is used in this work. GiBUU considers all the nucleons to be bound in a potential $U(r,p)$ as a function of coordinates and momentum. This potential can be obtained from the nuclear matter binding properties and analysis of p-A reactions \cite{Welke:1988zz}. The form of this potential is such that the nucleons having higher momentum experience a less attractive potential than those with lower momentum. GiBUU uses LFG to model the momentum distribution using $p_F \sim \rho^{1/3}$. The ground state of the nucleus is described by a realistic nuclear density profile and the nuclear potential $U(r,p)$ can be calculated using an energy-density functional. Finally, nucleons are placed into this potential with a momentum distribution according to the LFG model. The density profiles used in this simulation are Woods-Saxon density distribution considering different radii for nucleons, and a spherical distribution according to the input value of Fermi momentum. The nucleon removal energy is taken as 15 MeV. The QE scattering is simulated using the LS model considering the dipole form of the axial form factor with an axial mass of 1.03 GeV. GiBUU considers 13 resonance channels with the vector form factors calculated from the electron scattering data using the MAID analysis \cite{Drechsel:1998hk}. PYTHIA handles the DIS simulation considering an invariant mass of around 2 GeV. The treatment of FSI is based on quantum-kinetic transport theory \cite{kadanoff2018quantum}. The transport approach models the time evolution of hadrons using the Boltzmann-Uehling-Uhlenbeck (BUU) equation, which describes their interactions with nucleons and mesons. The mean-field potentials affect the motion of hadrons, altering their trajectories and energy distributions within the nucleus.

\section{\label{sec4}Results}

\begin{figure*}
\centering
\begin{subfigure}{0.49\textwidth}
    \centering
    \includegraphics[width=9cm,height=8.5cm]{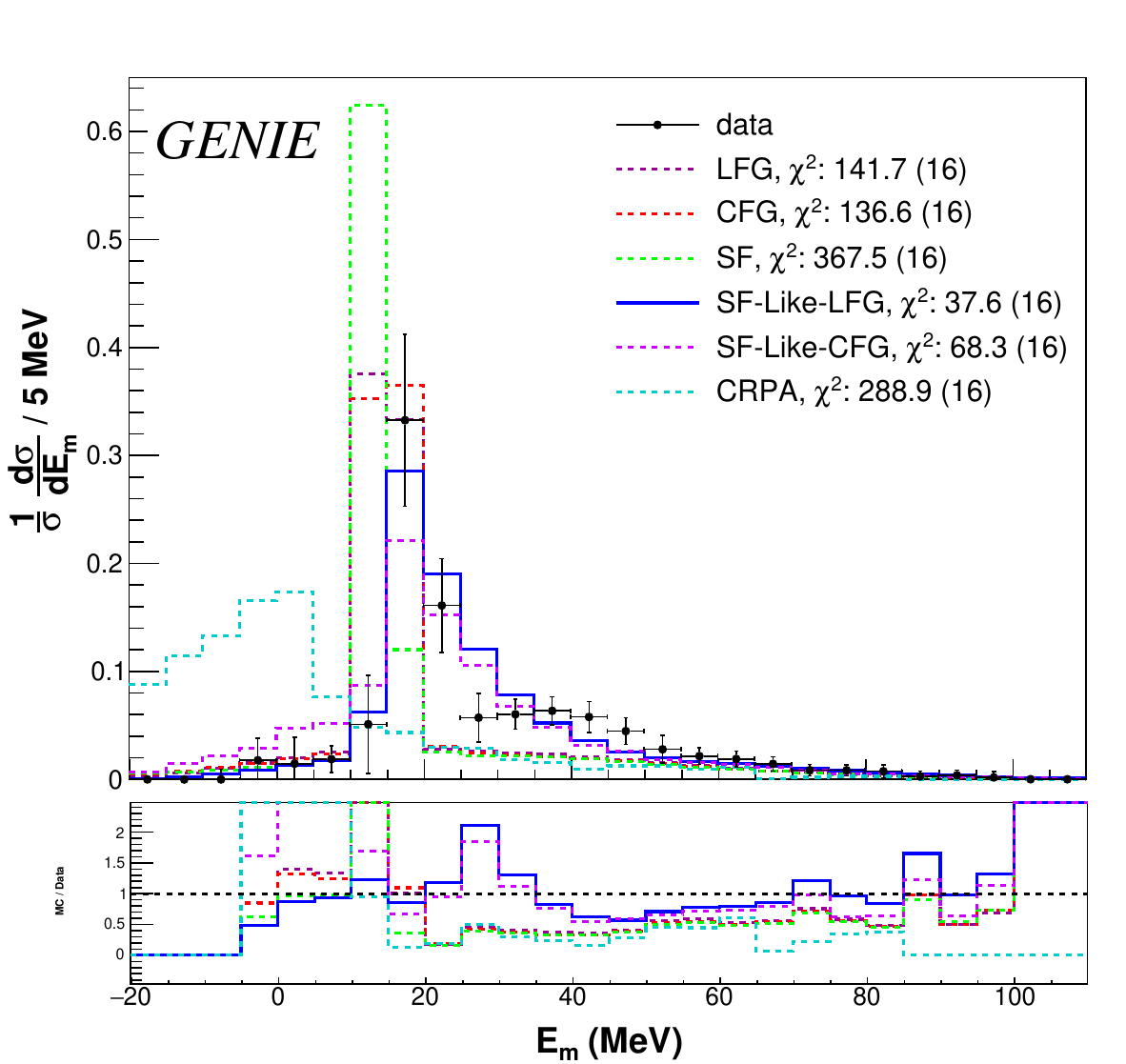}
    
\end{subfigure}
\hfill
\begin{subfigure}{0.49\textwidth}
    \centering
    \includegraphics[width=9cm,height=8.5cm]{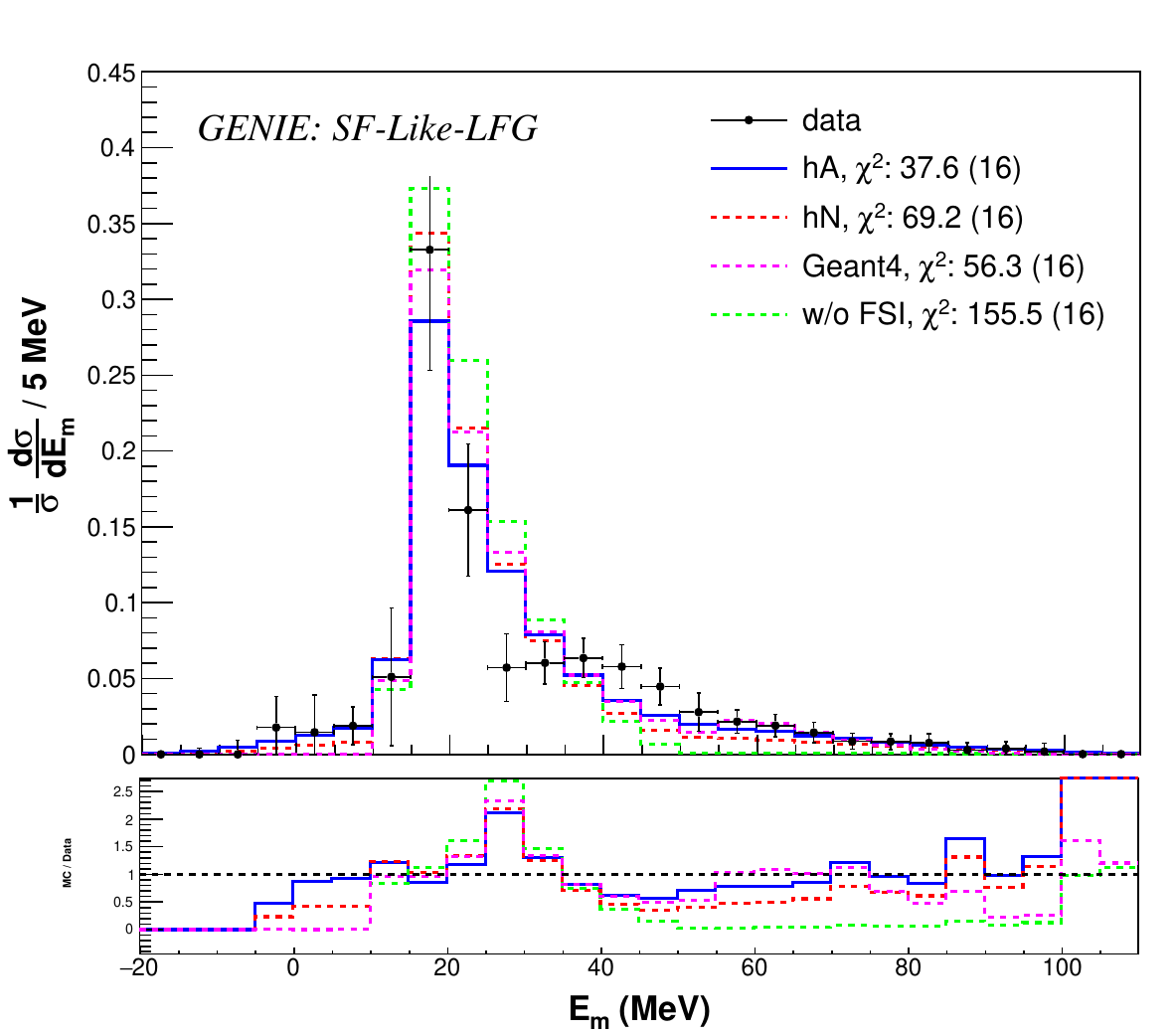}
    
\end{subfigure}
\caption{The shape-only differential cross-section in missing energy from GENIE using different nuclear (left) and FSI (right) models.}
\label{fig:1}
\end{figure*}

The missing energy is reconstructed using Eq. \ref{eq1} from the final-state particles of selected events with $1\mu^- 1p\,0\pi \,Xn$ in the final state. To quantify the comparison between the data ($d$) and Monte Carlo (MC) prediction ($p$), $\chi^2 = \sum_{ij} (d_i-p_i)V_{ij}^{-1}(d_j-p_j)$ is calculated using the covariance matrix $V_{ij}$. $\chi^2$ is calculated across 16 bins in the range of $E_m$ = 5 to 85 MeV to account for the higher-statistics region.

\subsection{GENIE}

The shape-only differential cross-section in missing energy using GENIE is shown in Fig.\ref{fig:1} for different nuclear models and FSI models. The left panel shows the predictions for different nuclear models. The peak around 17 MeV is only predicted by SF-like-LFG and SF-like-CFG. The peak is mainly due to the nucleon separation energy; however, LFG, CFG, and SF don't explain the peak around 17 MeV. SF-like models treat removal energy differently by considering it momentum-dependent. It changes the Q-value to be the binding energy per nucleon, and the removal energy will be the sum of the removal energy at the Fermi surface and the Q-value, minus the initial state nucleon kinetic energy. But LFG and CFG consider the removal energy to be the binding energy per nucleon. SF-like-CFG has an over-prediction compared to the SF-like-LFG, causing a higher $\chi^2$. The SF considers an effective removal energy for a range of A (total number of nucleons) values. The ratio plot between MC and data shows the ratio for LFG, CFG, and SF converges towards 1 in the higher missing energy region above 40 MeV. The prediction from the CRPA model shows a shifted distribution towards lower missing energy than the other nuclear model with a high $\chi^2$ = 288.9 (16). The cause of such a shift from the CRPA model at lower neutrino energies \cite{Dolan:2021rdd} will be studied in the future. The SF-like-LFG predicts better results in comparison to the other nuclear models with a $\chi^2$ = 37.6 (16). The ratio plot shows a higher deviation from 1 around 30 MeV for the SF-like-LFG model. The right panel of Fig.\ref{fig:1} shows the cross-section for different FSI models: hA, hN, Geant4, and without FSI effect using the nuclear model SF-like-LFG. The prediction from the hA model shows a better agreement than the hN and Geant4 models with a significantly lower $\chi^2$. The ratio plot shows that the hN model provides a ratio close to 1 and predicts a higher peak around 15-20 MeV compared to the hA model. The hN model tracks each hadron step by step as it propagates through the nuclear medium, overestimating rescattering in the nucleus and leading to excessive energy loss. In contrast, the hA model applies averaged probabilities, effectively representing nuclear transparency effects and preventing excessive missing energy \cite{Andreopoulos:2015wxa}. Geant4 has incorporated the ability to simulate low-energy compound nuclear reactions and coalescence processes, allowing for the production of light ions and photons in the final state. It models the nuclear structure using multiple concentric shells at various radii, each with its own specific depth. This approach enables the inclusion of medium effects and binding energy corrections in a simplified manner. The prediction from Geant4 shows no events in the lower missing energy region below $\sim$ 10 MeV. The prediction without FSI became narrower, and the event rate increased in the peak region, indicating that the broad prediction above $\sim$ 40 MeV and below $\sim$ 10 MeV is due to the FSI effect. The hA model with the SF-like-LFG nuclear model provides the best prediction of missing energy among the nuclear and FSI models in GENIE.

\subsection{GENIE Tuning}

From the right panel of Fig. \ref{fig:1}, it is observed that FSI effects have a significant impact on the cross section in missing energy. Since the hA model gives a better agreement, the fractions of hA FSI fates and the mean free path are considered as tuning parameters. The tuning process uses the GENIE reweighting package v1.0.4. The tweaking of the mean free path ($\lambda^h$) reweights the probability of rescattering of each hadron $h$ within the nuclear medium ($p_{rescat}^h$). The rescattering probability is calculated by \cite{Andreopoulos:2015wxa},
\begin{equation*}
    p_{rescat}^h = 1 - \int{e^{-r/\lambda^h(\vec{r},E_h)} dr}
\end{equation*}
$\lambda^h$ is a function of hadron's position ($\vec{r}$), and its energy ($E_h$) within the nucleus. $\lambda^h$ can be modified by the tweaking dial ($S_{\lambda}^h$) with corresponding error $\delta \lambda^h$ as follow:
\begin{equation*}
    \lambda^h \rightarrow \lambda^{h'} = \lambda^h(1+S_{\lambda}^h \times \frac{\delta\lambda^h}{\lambda^h})
\end{equation*}

\begin{figure*}
\centering
\begin{subfigure}{0.49\textwidth}
    \centering
    \includegraphics[width=9cm,height=7cm]{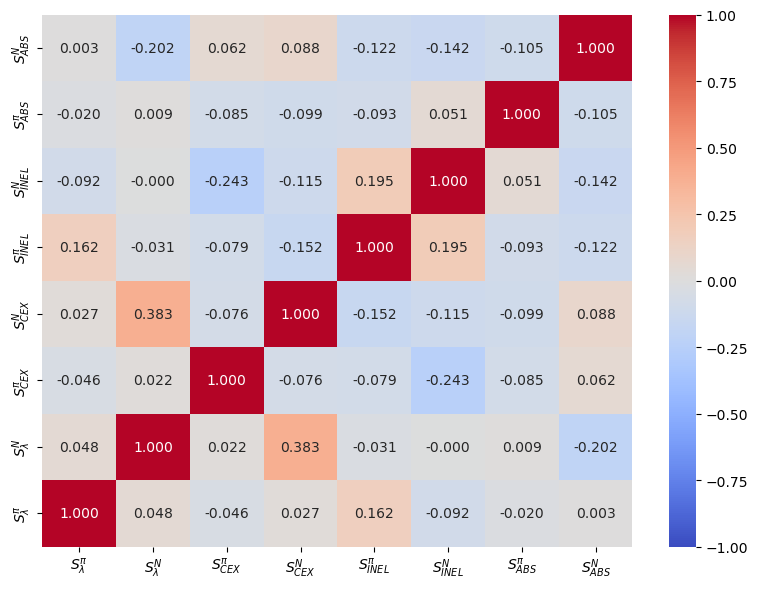}
\end{subfigure}
\hfill
\begin{subfigure}{0.49\textwidth}
    \centering
    \includegraphics[width=9cm,height=7cm]{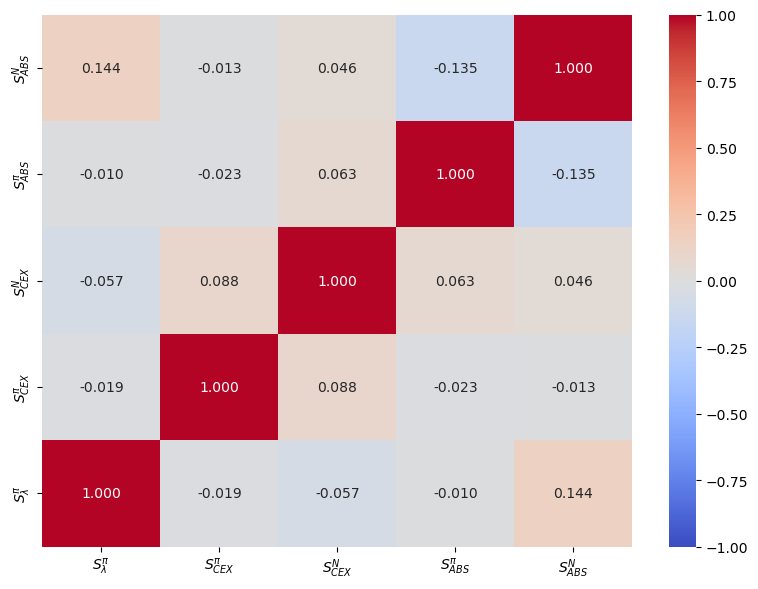}
\end{subfigure}
\caption{Post-fit correlation between FSI fates for all parameters (left) and selected parameters (right).}
\label{fig:correlation}
\end{figure*}

\begin{table*}[]
\centering
\setlength{\tabcolsep}{10pt}
\renewcommand{\arraystretch}{1.8}
\begin{tabular}{|cccccc|}
\hline
Parameter                 & Nominal              & Range      & All-parameter (Best-fit)             & Selected-parameter (Best-fit)        & Final tune \\ \hline
    $S_{\lambda}^{\pi}$   &    1.0 $\pm$ 0.2     & (0.0, 3.0) & 1.24 $\pm$ 0.56    & 1.12 $\pm$ 0.53 & \Checkmark         \\
    $S_{\lambda}^{N}$     &    1.0 $\pm$ 0.2     & (0.0, 3.0) & 0.03 $\pm$ 0.32    & 1.0             &            \\
    $S_{CEX}^{\pi}$       &    1.0 $\pm$ 0.5     & (0.0, 3.0) & 2.03 $\pm$ 0.50    & 0.58 $\pm$ 0.52 & \Checkmark           \\
    $S_{CEX}^{N}$         &    1.0 $\pm$ 0.5     & (0.0, 3.0) & 1.57 $\pm$ 0.44    & 2.74 $\pm$ 0.46 & \Checkmark           \\
    $S_{INEL}^{\pi}$      &    1.0 $\pm$ 0.4     & (0.0, 3.0) & 1.38 $\pm$ 0.53    & 1.0             &            \\
    $S_{INEL}^{N}$        &    1.0 $\pm$ 0.4     & (0.0, 3.0) & 1.31 $\pm$ 0.43    & 1.0             &            \\
    $S_{ABS}^{\pi}$       &    1.0 $\pm$ 0.2     & (0.0, 3.0) & 0.86 $\pm$ 0.47    & 1.60 $\pm$ 0.48 & \Checkmark           \\
    $S_{ABS}^{N}$         &    1.0 $\pm$ 0.2     & (0.0, 3.0) & 2.07 $\pm$ 0.55    & 0.62 $\pm$ 0.53 & \Checkmark           \\ \hline
\end{tabular}
\caption{Tunable parameters of the hA model and their best-fit values after tuning using all parameters and selected parameters.}
\label{tab:1}
\end{table*}

\begin{figure}[!h]
    \centering
    \includegraphics[width=9cm,height=8.5cm]{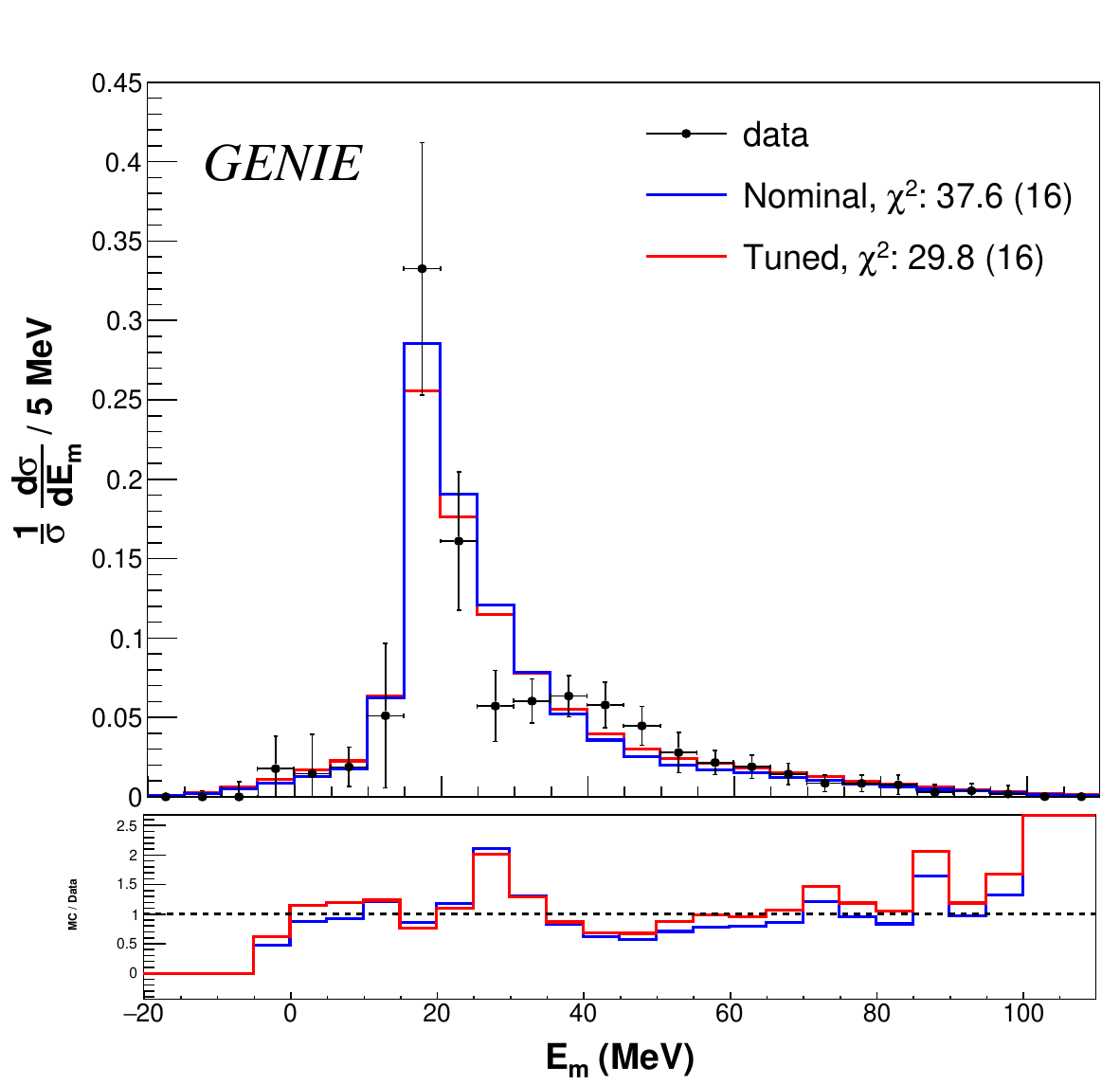}
    \caption{The shape-only differential cross-section in missing energy for nominal and tuned parameters using GENIE.}
    \label{fig:5}
\end{figure}

The weight corresponding to the change in $\lambda^h$ can be calculated from the nominal ($p_{rescat}^h$) and tweaked ($p_{rescat}^{h'}$) rescattering probability \cite{Andreopoulos:2015wxa}, 
\begin{equation*}
    w_{\lambda}^h = 
    \begin{cases}
    \frac{p_{rescat}^{h'}}{p_{rescat}^{h}} & \text{if } h \text{ re-interacts} \\
    \\
    \frac{1-p_{rescat}^{h'}}{1-p_{rescat}^{h}} & \text{if } h \text{ escapes}
    \end{cases}
\end{equation*}

Change in FSI hadron fates modifies the hadron-nucleus cross section for a particular fate ($f$), with the corresponding error in the cross section ($\delta \sigma_f^h$) as in:

\begin{equation*}
    \sigma_f^h \rightarrow \sigma_f^{h'} = \sigma_f^h(1+S_{f}^h \times \frac{\delta \sigma_f^h}{\sigma_f^h})
\end{equation*}
where $S_f^h$ is the fate tweaking dial. The weight corresponding to the fate tweaking parameter is calculated as:
\begin{equation*}
    w_f^h = \sum_f{\delta_{ff'} \times S_f^h \times \frac{\delta \sigma_f^h}{\sigma_f^h} }
\end{equation*}

$f$ runs over all the possible fates, such as charge exchange, inelastic, absorptions, and pion production. $f'$ is the actual fate of the hadron in the simulation. Finally, the event weight can be calculated by,
\begin{equation*}
    w^{evt} = \prod_j{(w_{\lambda}^h \times w_{f}^h)_j}
\end{equation*}
where $j$ runs over all the particles in the primary hadronic system in a particular event.

The sum of all fate fractions must equal 1 to conserve total probability. Therefore, at least one fate cannot be directly tweaked—this is known as the cushion term. It is automatically adjusted to absorb changes in the other fate fractions. Since the work analyses the final states with 0 pions, the pion production fates are considered as the cushion terms. Due to the complex correlations among model parameters, identifying those most constrained by the data is challenging. Significant correlations and anti-correlations are observed between different FSI fates shown in Fig. \ref{fig:correlation}. The tuning began by considering all parameters with values in the range 0 to 3, with nominal values and priors \cite{Andreopoulos:2015wxa} as given in Table \ref{tab:1}. The best-fit combination of all parameters shows that some parameters are close to their nominal values, within 1$\sigma$ of the prior. These parameters are removed from the tuning, and their exclusion does not significantly impact the result. The mean free path of nucleons ($\lambda_N$) is also excluded from the tuning. A reduction in $\lambda_N$ increases the rescattering of nucleons inside the nucleus, leading to a decrease in cross section in $E_m$ around the peak region of $E_m\approx 17$ MeV and vice versa. The best fit combination of the selected parameters ($S_{\lambda}^{\pi},\,S_{CEX}^{\pi},\,S_{CEX}^{N},\,S_{ABS}^{\pi},\,S_{ABS}^{N}$) is given in Table \ref{tab:1} and the $\chi^2$ for cross section with the tuned parameters is significantly reduced to 29.8 (16) shown in Fig. \ref{fig:5}. The ratio plot between MC and data shows an improvement for the tuned distribution, with the ratio being close to 1 compared to the nominal distribution. However, the dip around 25 MeV remained unexplained. The tuning of the hA FSI model improved the cross section, but it is not enough to improve the shape of the $E_m$ distribution.

\subsection{NuWro}

\begin{figure}[!h]
    \centering
    \includegraphics[width=9cm,height=8.5cm]{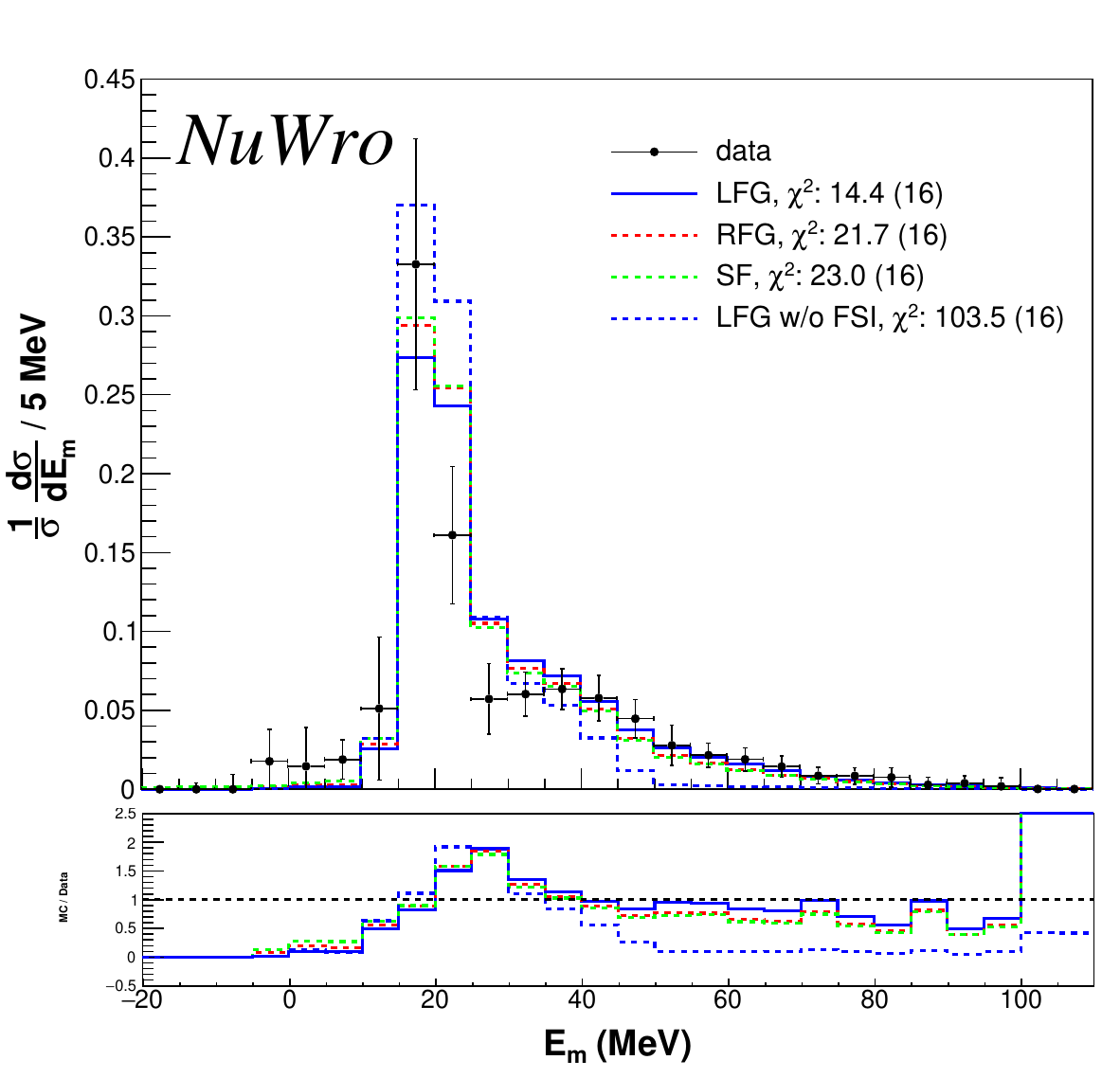}
    \caption{The shape-only differential cross-section in missing energy from NuWro using different nuclear models.}
    \label{fig:2}
\end{figure}

\begin{figure*}
\centering
\begin{subfigure}{0.49\textwidth}
    \centering
    \includegraphics[width=9cm,height=7.2cm]{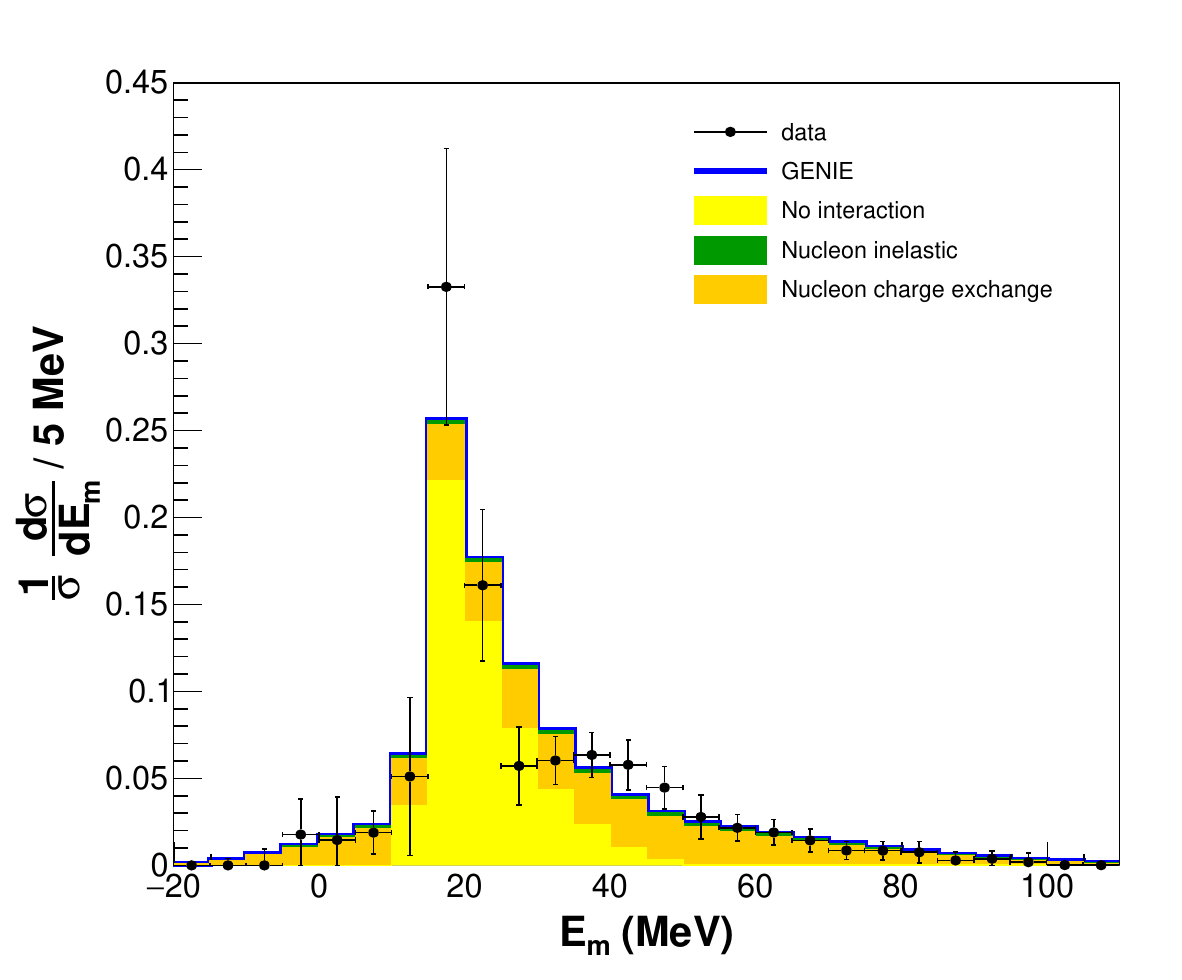}
\end{subfigure}
\hfill
\begin{subfigure}{0.49\textwidth}
    \centering
    \includegraphics[width=9cm,height=7.2cm]{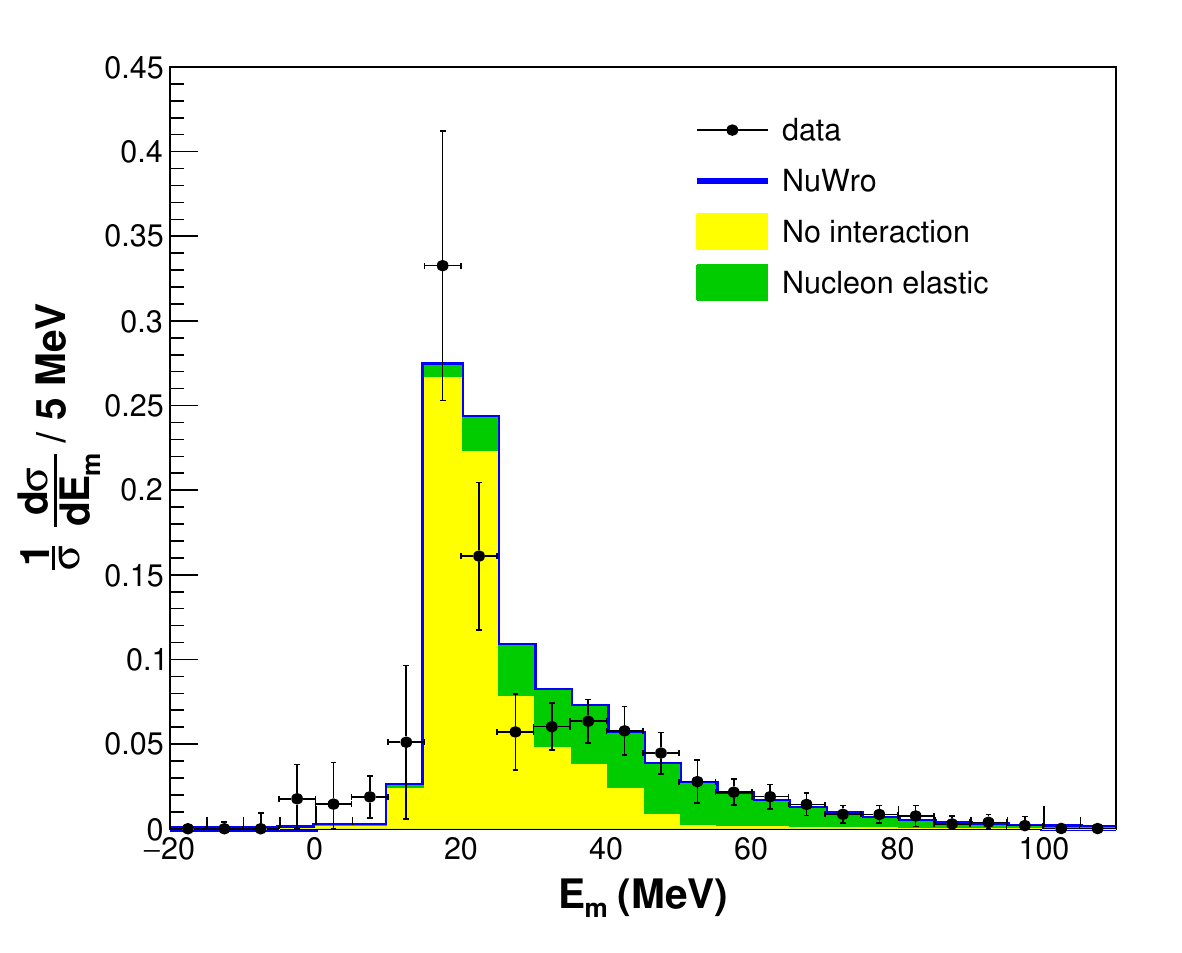}
\end{subfigure}
\caption{Contributions from different FSI fates to the missing energy for GENIE (left) and NuWro (right).}
\label{fig:fsi_fates}
\end{figure*}

The differential cross-sections in reconstructed missing energy, using the nuclear models (LFG, RFG, and SF) in NuWro, are shown in Fig. \ref{fig:2}. All three models predict similar shapes for the missing energy spectrum and show the peak around 17 MeV. LFG and RFG treat removal energy similarly, assuming constant binding energy, whereas SF employs a realistic distribution of removal energy. However, differences in the treatment of removal energy among these models do not significantly impact the reconstructed missing energy. The predictions overestimate the data in lower missing energy regions below 15 MeV for all three nuclear models. SF predicts a higher event rate at the peak than the other models due to multi-nucleon knockout effects since it considers the short-range correlations between nucleons. The ratio plot between MC and data shows the LFG model provides the ratio closer to unity in the region of missing energy above 30 MeV than the other models. NuWro shows a similar effect as GENIE without FSI: the event rate increased in the peak region and the prediction became narrow. The ratio also remains close to 0 above $\sim$ 50 MeV when the FSI effect is disabled. This similarity to GENIE without FSI suggests that NuWro’s improved agreement with data may point to qualitative differences in its modeling of FSI. Figure \ref{fig:fsi_fates} shows that there is a significant contribution from nucleon elastic scattering in NuWro, which is replaced with the no-FSI component in the GENIE hA FSI model to explain the MINER$\nu$A data \cite{MINERvA:2023avz, Harewood:2019rzy}. The nucleon elastic FSI could be one of the reasons behind the improved prediction in NuWro, in contrast to GENIE. Overall, the prediction from the LFG model provides the best fit to the data, with $\chi^2 = 14.4$ (16), compared to the other nuclear models in NuWro.

\subsection{GiBUU}

\begin{figure}[!h]
    \centering
    \includegraphics[width=9cm,height=8.5cm]{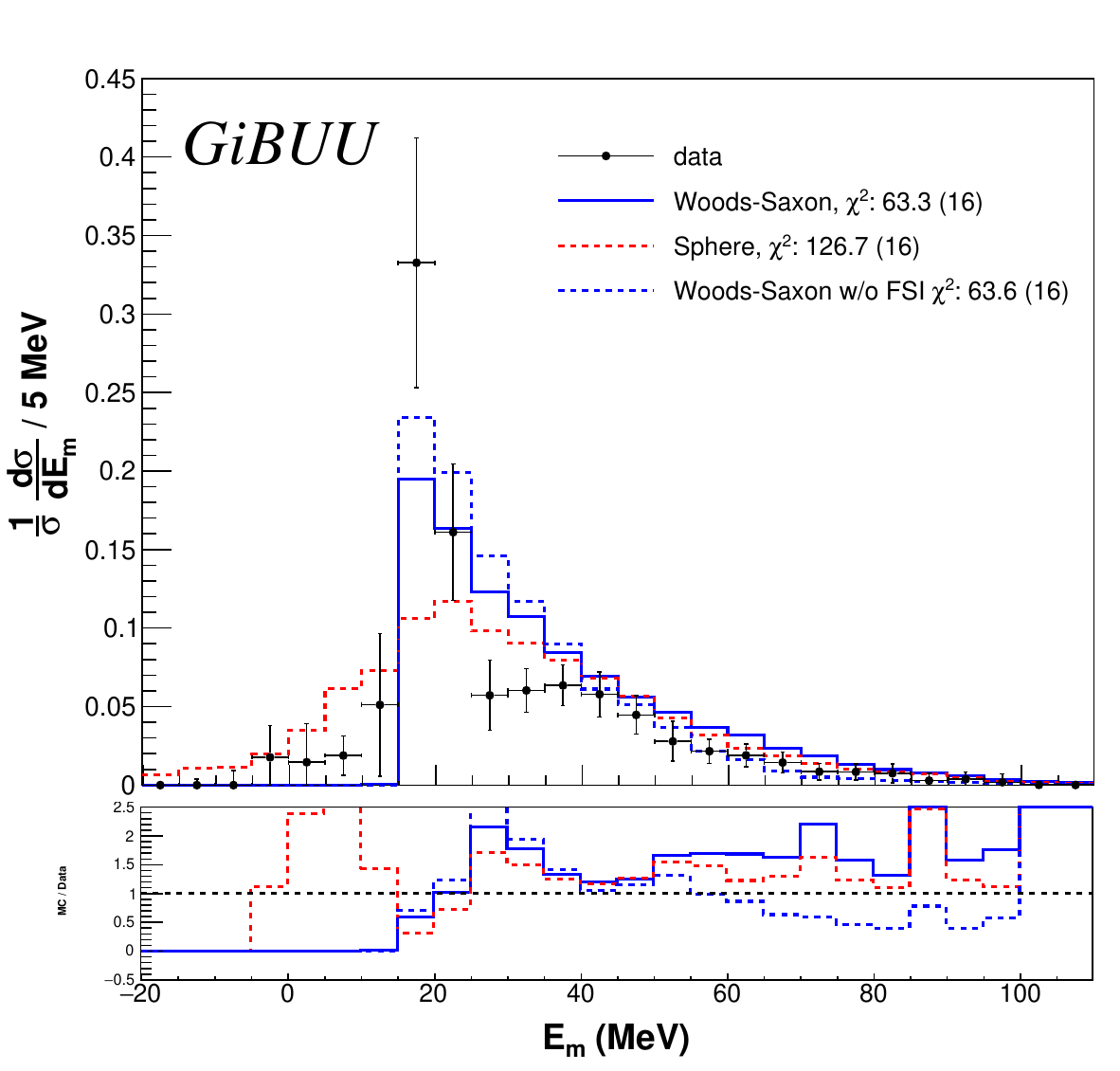}
    \caption{Same as Fig.\ref{fig:2} using GiBUU.}
    \label{fig:3}
\end{figure}

Fig.\ref{fig:3} shows GiBUU predictions for the missing energy distribution using two nuclear density profiles: the Woods-Saxon distribution and a spherical distribution. The shape of the differential cross-section differs between the two nuclear density distributions.  GiBUU does not explain the peak around 17 MeV; however, the Woods-Saxon model exhibits a small peak at this energy due to the constant nucleon separation energy of 15 MeV. It predicts a broad missing energy spectrum with no events below $\sim$10 MeV. The issue of the broader spectrum is discussed in Ref. \cite{Gallmeister:2025iug}, which investigates the effects of binding energy corrections, in-medium collision, and in-medium propagation. The ratio plot shows that the Woods-Saxon model underestimates the data in the missing energy region above 25 MeV. The spherical density profile produces a similarly broad spectrum in this energy range; however, its ratio plot is closer to 1 compared to that of the Woods-Saxon model. Below 15 MeV, the spherical density underestimates the data, leading to a high $\chi^2$ of 126.7 (16). In the low-missing-energy region (below $\sim$10 MeV), the Woods-Saxon model predicts no events similar to the GENIE predictions with the Geant4 FSI model, whereas the spherical density underestimates the data. This suggests that the Woods-Saxon model requires modification such that it behaves similarly to spherical density in that energy range. To check the effect of FSI in GiBUU, the missing energy spectrum was reconstructed with FSI disabled. This results in an increased event rate around the peak region and a decreased event rate at higher missing energy (above $\sim$60 MeV). However, overall, FSI does not significantly impact the missing energy spectrum for the Woods-Saxon density, yielding a $\chi^2$ of 63.6 (16). The Woods-Saxon model provides a better overall agreement with the data, as indicated by its lower $\chi^2$ of 63.3 (16) compared to the spherical distribution.

\begin{figure}[!h]
    \centering
    \includegraphics[width=9cm,height=8.5cm]{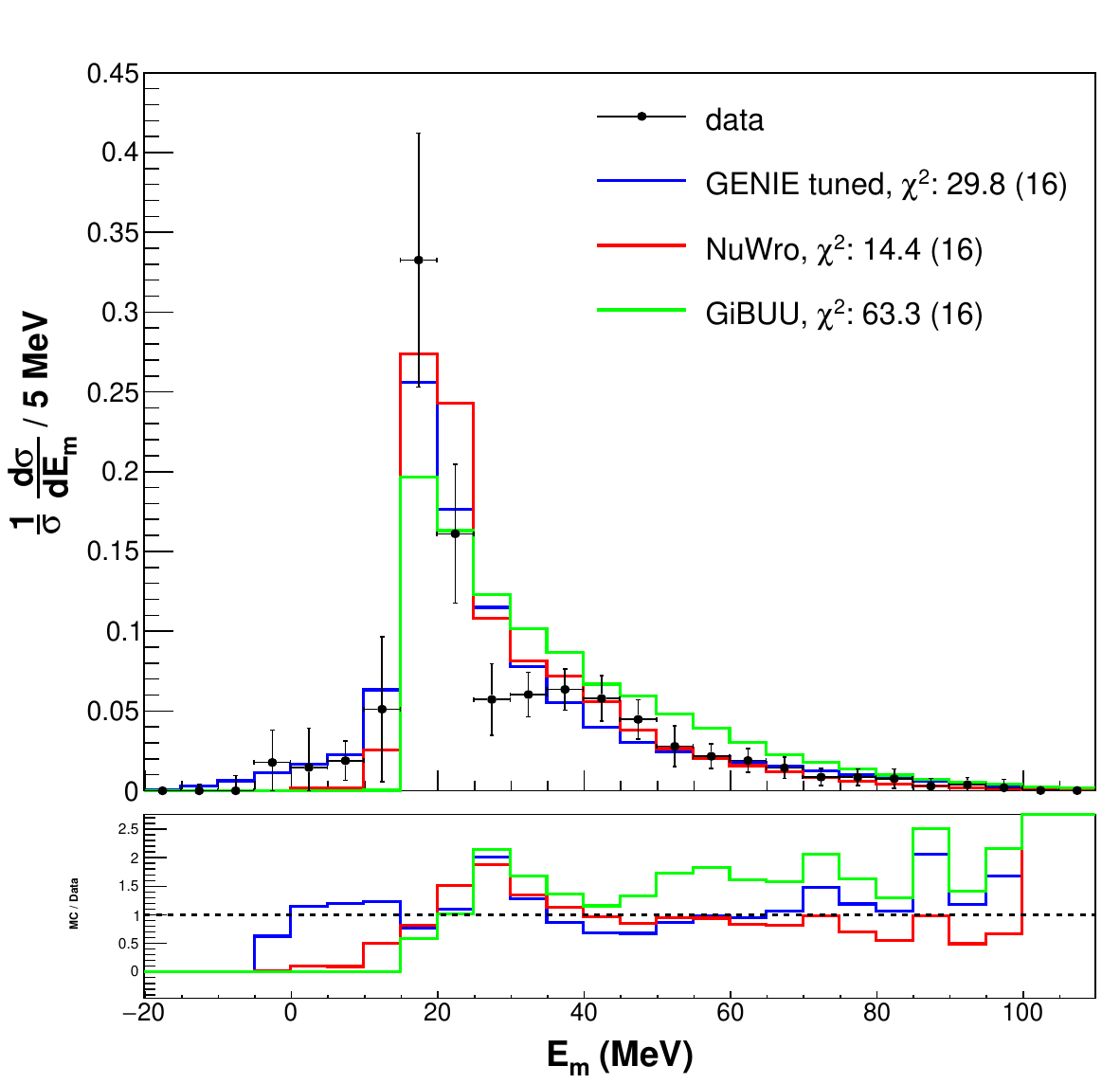}
    \caption{The shape-only differential cross-section in reconstructed missing energy using GENIE, NuWro, and GiBUU.}
    \label{fig:4}
\end{figure}

\section{\label{sec5}Conclusion}

In this work, we explore the effect of nuclear models and FSI models on the missing energy in KDAR $\nu_\mu$ CC scattering on carbon using JSNS$^2$ measurement. NuWro provides an overall better fit as shown in Fig.\ref{fig:4}. The ratio plot shows NuWro provides a better prediction in the region of missing energy $\sim$ 30 - 70 MeV than GENIE tuned; however, GENIE tuned is better in the lower missing energy below $\sim$ 20 MeV. The contributions from FSI components suggest that the nucleon elastic channel could be the factor that improves the description of data in NuWro. On the other hand, GiBUU shows a broader spectrum. The nucleon removal energy causes the peak at 15 MeV observed in GENIE and GiBUU. The predictions from all three MC generators underestimate the data in the region,  $E_m\sim$ 20 - 30 MeV, which should be investigated in future studies. The SF-like-LFG in GENIE, the LFG in NuWro, and the Woods-Saxon model in GiBUU provide the best-fit predictions for each generator. This indicates that different treatments of the ground state of the nucleus in generators require improvement. The nuclear and FSI models notably affect the reconstructed missing energy. However, the substantial discrepancy between data and MC could arise from various nuclear effects, such as Fermi motion, RPA, SRC, and Pauli blocking. Further studies on these nuclear effects are necessary for a better explanation of the data. Since KDAR neutrinos have a known neutrino energy, they provide an opportunity to reduce the uncertainties and to improve the understanding of low-energy neutrino interactions.

\section*{Acknowledgements} 
R K Pradhan acknowledges the DST-INSPIRE grant (2022/IF220293) for financial support. A Giri credited the grant support of the Department of Science and Technology (SR/MF/PS-01/2016-IITH/G). We thank Dr. Stephen Dolan for useful discussions on the CRPA model.

\bibliography{references}

\end{document}